\documentclass[sigconf]{acmart}

\usepackage{amsmath,amssymb,amsfonts}
\usepackage{algorithmic}
\usepackage{graphicx}
\usepackage{textcomp}
\usepackage{hyperref}
\usepackage{tikz}
\usepackage{xcolor}
\usepackage{tabularx}
\usepackage{multicol}
\usepackage{multirow}
\usepackage{booktabs} % For formal tables
\usepackage{color}
\usepackage[utf8]{inputenc}
\usepackage{hyperref}

\usepackage{listings}
\usepackage{color}
\usepackage[utf8]{inputenc}
\usepackage[T1]{fontenc}
\usepackage{microtype}
\usepackage{amsmath}

\newcommand{\fakeparagraph}[1]{\smallskip\noindent\textbf{#1.}}
\newcommand*\circled[1]{\tikz[baseline=(char.base)]{
            \node[shape=circle,draw,inner sep=1pt] (char) {#1};}}
\newcommand{\nextitem}{\par\hspace*{\labelsep}\textbullet\hspace*{\labelsep}}

\AtBeginDocument{%
  \providecommand\BibTeX{{%
    \normalfont B\kern-0.5em{\scshape i\kern-0.25em b}\kern-0.8em\TeX}}}

\begin{document}

\copyrightyear{2020}
\acmYear{2020}
\setcopyright{acmlicensed}\acmConference[IoT '20]{10th International Conference on the Internet of Things}{October 6--9, 2020}{Malmö, Sweden}
\acmBooktitle{10th International Conference on the Internet of Things (IoT '20), October 6--9, 2020, Malmö, Sweden}
\acmPrice{15.00}
\acmDOI{10.1145/3410992.3411013}
\acmISBN{978-1-4503-8758-3/20/10}

%%
%% The "title" command has an optional parameter,
%% allowing the author to define a "short title" to be used in page headers.
%

\title{A Distributed Framework to Orchestrate Video Analytics Applications}
%%
%% The "author" command and its associated commands are used to define
%% the authors and their affiliations.
%% Of note is the shared affiliation of the first two authors, and the
%% "authornote" and "authornotemark" commands
%% used to denote shared contribution to the research.

\author{Tapan Pathak, Vatsal Patel, Sarth Kanani, Shailesh Arya}
%\authornote{All authors contributed equally to this research.}
%\email{tapan.pce18,vatsal.pce18, sarth.kce18, shailesh.ace16@sot.pdpu.ac.in}
% \orcid{1234-5678-9012}
%\author{G.K.M. Tobin}
%\authornotemark[1]
%\email{webmaster@marysville-ohio.com}
\affiliation{%
  \institution{Pandit Deendayal Petroleum University}
  %\streetaddress{P.O. Box 1212}
  \city{Gandhinagar}
  %\state{Gujarat}
  %\postcode{43017-6221}
  \country{India}
}

\author{Pankesh Patel}
%\authornote{All authors contributed equally to this research.}
\email{pankesh.patel@insight-centre.org}
%\email{ali.intizar@nuigalway.ie}
% \orcid{1234-5678-9012}
%\author{G.K.M. Tobin}
%\authornotemark[1]
%\email{webmaster@marysville-ohio.com}
\affiliation{%
  \institution{SFI Confirm Centre for Smart Manufacturing, Data Science Institute, NUI Galway, Ireland}
  %\streetaddress{P.O. Box 1212}
  %\city{Gandhinagar}
  %\state{Ohio}
  %\postcode{43017-6221}
}

\author{Muhammad Intizar Ali}
%\authornote{All authors contributed equally to this research.}
%\email{pankesh.patel@insight-centre.org}
\email{ali.intizar@nuigalway.ie}
% \orcid{1234-5678-9012}
%\author{G.K.M. Tobin}
%\authornotemark[1]
%\email{webmaster@marysville-ohio.com}
\affiliation{%
  \institution{SFI Confirm Centre for Smart Manufacturing, Data Science Institute, NUI Galway, Ireland}
  %\streetaddress{P.O. Box 1212}
  %\city{Gandhinagar}
  %\state{Ohio}
  %\postcode{43017-6221}
}

\author{John Breslin}
%\authornote{All authors contributed equally to this research.}
\email{john.breslin@nuigalway.ie}
%\email{ali.intizar@nuigalway.ie}
% \orcid{1234-5678-9012}
%\author{G.K.M. Tobin}
%\authornotemark[1]
%\email{webmaster@marysville-ohio.com}
\affiliation{%
  \institution{SFI Confirm Centre for Smart Manufacturing, Data Science Institute, NUI Galway, Ireland}
  %\streetaddress{P.O. Box 1212}
  %\city{Gandhinagar}
  %\state{Ohio}
  %\postcode{43017-6221}
}

%%
%% By default, the full list of authors will be used in the page
%% headers. Often, this list is too long, and will overlap
%% other information printed in the page headers. This command allows
%% the author to define a more concise list
%% of authors' names for this purpose.
\renewcommand{\shortauthors}{Pathak and Patel, et al.}

%%
%% The abstract is a short summary of the work to be presented in the
%% article.
\begin{abstract}
The concept of Internet of Things~(IoT) is a reality now. This paradigm shift has caught everyone’s attention in a large class of applications, including IoT-based video analytics using smart doorbells. The global market for IoT-based video analytics is expected to grow to 9.4 billion USD by 2025. Due to its growing application segments, various efforts exist in scientific literature and many video-based doorbell solutions are commercially available in the market. However, contemporary offerings are bespoke, offering limited composability and reusability of a smart doorbell framework. Second, they are monolithic and proprietary, which means that the implementation details remain hidden from the users. We believe that a transparent design can greatly aid in the development of a smart doorbell, enabling its use in multiple application domains.

To address the above-mentioned challenges, we propose a distributed framework to orchestrate video analytics across Edge and Cloud resources. We investigate trade-offs in the distribution of different software components over a bespoke/full system, where components over Edge and Cloud are treated generically. The proposal looks at the smart doorbell case study and uses AWS as a base platform for implementation. To showcase the composability and reusability of the proposed framework, the state of the art ML-and DL-based models are implemented on the affordable Raspberry Pi in the form of an Edge device. This also demonstrates the feasibility of running these models on IoT devices. Moreover, this paper evaluates the proposed framework as well as the state-of-the-art models and presents comparative analysis of them on various metrics~(such as overall model accuracy, latency, memory and CPU usage). The evaluation result demonstrates our intuition very well, showcasing that the AWS-based approach exhibits reasonably high object-detection accuracy, low memory and CPU usage when compared to the state-of-the-art approaches, but high latency.

\end{abstract}



%%
%% The code below is generated by the tool at http://dl.acm.org/ccs.cfm.
%% Please copy and paste the code instead of the example below.
%%
% \begin{CCSXML}
% <ccs2012>
%  <concept>
%   <concept_id>10010520.10010553.10010562</concept_id>
%   <concept_desc>Computer systems organization~Embedded systems</concept_desc>
%   <concept_significance>500</concept_significance>
%  </concept>
%  <concept>
%   <concept_id>10010520.10010575.10010755</concept_id>
%   <concept_desc>Computer systems organization~Redundancy</concept_desc>
%   <concept_significance>300</concept_significance>
%  </concept>
%  <concept>
%   <concept_id>10010520.10010553.10010554</concept_id>
%   <concept_desc>Computer systems organization~Robotics</concept_desc>
%   <concept_significance>100</concept_significance>
%  </concept>
%  <concept>
%   <concept_id>10003033.10003083.10003095</concept_id>
%   <concept_desc>Networks~Network reliability</concept_desc>
%   <concept_significance>100</concept_significance>
%  </concept>
% </ccs2012>
% \end{CCSXML}

% \ccsdesc[500]{Computer systems organization~Embedded systems}
% \ccsdesc[300]{Computer systems organization~Redundancy}
% \ccsdesc{Computer systems organization~Robotics}
% \ccsdesc[100]{Networks~Network reliability}

%%
%% Keywords. The author(s) should pick words that accurately describe
%% the work being presented. Separate the keywords with commas.

%% A "teaser" image appears between the author and affiliation
%% information and the body of the document, and typically spans the
%% page.
% \begin{teaserfigure}
%   \includegraphics[width=\textwidth]{sampleteaser}
%   \caption{Seattle Mariners at Spring Training, 2010.}
%   \Description{Enjoying the baseball game from the third-base
%   seats. Ichiro Suzuki preparing to bat.}
%   \label{fig:teaser}
% \end{teaserfigure}

%%
%% This command processes the author and affiliation and title
%% information and builds the first part of the formatted document.
\maketitle

\section{Introduction}\label{sec:intro}

The Internet of Things (IoT)-based video analytics has caught the attention of businesses and technology enthusiasts worldwide, due to its potential to provide solutions in a large class of applications including Smart-X~(e.g., Smart home, Smart farming). For example, a video-based smart doorbell announces the presence of visitors. This would help the homeowners to verify the identity of visitors, which could prevent unwanted invasion or home robbery at a moment of notice~\cite{low-cost-doorbell}. A video-based smart doorbell may leverage face-detection technologies to measure unique characteristics about one's face. Due to its growing application segments, various efforts exist in literature~\cite{low-cost-doorbell,iris-doorbell, iot-based-smartdoorbell} and many video-based doorbell solutions are commercially available in the market~\cite{Delaney2020}. Some of the limitations of these offerings that we noted are:

Existing smart doorbell devices often have limited processing capabilities due to either their portability requirements or to lower the cost of installation, deployment and maintenance. On the other hand, the rise of DL/CV algorithms~(to improve the object-detection accuracy at the Edge) demands powerful but expensive devices~(e.g., NVIDIA Jetson Nano~\cite{nvidia2020}, Google Coral USB Accelerator~\cite{Google2020}). \textit{To address this trade-off, we exploit a distributed architecture that can orchestrate the placement of different components while keeping the trade-off between cost and detection accuracy in mind. Moreover, our experimental analyses will be informative to users~(e.g., software architects, system designers, developers) designing similar video analytics architectures. Moreover, our case study on smart video-based doorbell will be insightful to extend it to other design tasks.}

% Existing smart doorbell devices often have limited processing capabilities due to either their portability requirements or to lower the cost of installation, deployment and maintenance. On the other hand, the rise of DL/CV algorithms~(to improve the object-detection accuracy at the Edge) demands powerful but expensive devices~(e.g., NVIDIA Jetson Nano~\cite{nvidia2020}, Google Coral USB Accelerator~\cite{Google2020}). To address this trade-off, \textit{a system design should exploit a distributed architecture that can orchestrate the placement of different components while keeping the trade-off between cost and detection accuracy in mind. Our approach to investigating these trade-offs will be informative to users (e.g., software architects, system designers, developers) designing similar video analytics architectures, and our case study on smart video-based doorbell will be insightful to extend to other design tasks}.

Existing systems are bespoke and proprietary, which offer limited composability and reusability of a smart doorbell framework~\cite{scalable-framework}. This would increase the effort and time to incorporate the rapid advancements in DL-based computer vision approaches and new innovative features. For example, the technical users may want to build new DL-based models to improve the accuracy or to develop new innovative applications (e.g., to prevent package theft~\cite{CNBC2020}, to detect suspicious activities at the doorbell such as gun detection~\cite{gundetection}), apart from the offered standard features. To address such evolving requirements, a transparent system design is needed that offers the flexibility to deploy new innovative DL/ML models and features over time, reusing the existing smart doorbell system architecture. We believe that a transparent design can greatly aid the use of a smart doorbell in multiple application domains.

\fakeparagraph{Contributions} To address the aforementioned challenges and design goals, we propose a distributed framework for video analytics across Edge and Cloud resources. The contributions of this paper are as follows:

\fakeparagraph{Multi-layer design and implementation} A multi-layer design to manage and deploy video analytics applications across Edge and Cloud~(presented in Section~\ref{sec:sysdesign}). The proposal is based on state-of-the-art technology and uses AWS as a base platform for the implementation. To the best of our knowledge, the implementation design is the first that presents a detailed design, orchestrating Edge and Cloud components and their detailed evaluations on various metrics. We believe that the transparent design will greatly aid the development of a smart doorbell and bring together the different disciplines (such as Edge analytics, AI, Internet of Things, Computer vision) to identify the major challenges and research directions.

\fakeparagraph{Validation study and evaluation results} We conduct an extensive experimental validation study to demonstrate composability and reusability of the proposed framework, aimed at the smart doorbell case study~(discussed in Section~\ref{sec:doorbell}).  We validate the proposed framework~(presented in Section~\ref{sec:sysdesign}) as well as the state-of-the-art models (presented in Section~\ref{sec:sota}) and conduct comparative analysis of them on various metrics~(presented in Section~\ref{sec:evaluation}).

\fakeparagraph{Outline} The rest of the paper is structured as follows: The smart doorbell case study is presented in Section~\ref{sec:doorbell}. Section~\ref{sec:sota} presents the approaches that are state of the art and designed around the main frameworks for video analytics on Edge devices. In Section~\ref{sec:sysdesign}, we present a system design of the proposed approach and its implementation. Section~\ref{sec:evaluation} presents the evaluation results of the proposed approach and its comparative analysis of the state-of-the-art approaches, presented in Section~\ref{sec:sota}. Finally, Section~\ref{sec:conclusion} concludes our current work and presents the future outlook.

\section{Case Study: Smart Doorbell}\label{sec:doorbell}
A homeowner deploys a small doorbell in his house, which provides personalized services for each family member. The smart doorbell comes with a mobile interface with an admin panel which allows each member to upload an image of each family member to train the face recognition module. The admin panel allows the user to add faces with friend, family and visitor categories. The face recognition modules categorize the faces according to the designated categories. If a new member and its categories are added, the smart doorbell updates the face recognition repository. The faces outside of these inputs are designated by the face recognition module as unknown. The deployed smart doorbell recognizes faces of family members, friends and visitors, and notifies the user in real time on his mobile app. In addition to this, it stores the past video footage of any activity in the proximity of the doorbell in order to analyze it more closely. Moreover, the doorbell records all past notifications. Also, the smart doorbell mobile app allows the users to watch a live view remotely by simply clicking on a view-live functionality.  

Apart from face recognition personalized services, the smart doorbell lets the homeowners develop innovative applications, leveraging the existing smart doorbell device and platform. For instance, it can be leveraged to detect \textbf{unsafe content}, which recognizes a weapon such as a gun, as it is the foremost weapon used for diverse crimes such as burglary, rape, etc~\cite{gundetection}. It can be an important technology to prevent package theft, as it is at an all-time high in recent years with 1.7 million packages stolen in the United States every day~\cite{CNBC2020}, thus making it \textbf{a noteworthy vehicle detection} functionality~(such as FedEx, USPS, etc.) at home. In addition to this, it provides advanced functions such as the detection of \textbf{known/unknown face} and \textbf{animal detection} at the door for the safety of homeowners.

In addition to the above-mentioned scenario, the user would like to add flexibility to add more devices on the fly with minimal effort. For instance, whenever a new smart doorbell is added to the home, the user connects the smart doorbell with the trained models of face recognition by simply configuring a new smart doorbell device, instead of training a new device again.

The next section presents the state-of-the-art approaches to implement the smart doorbell case study, demonstrating video analytics on IoT devices.

\section{State of the art}\label{sec:sota}
%  Many research efforts have focused on ways to reduce the latency, while improving object recognition accuracy, when it is executed on resource-constrained devices. Such efforts can have benefits throughout the ecosystem, by reducing the latency  while running on the end devices or edge servers. Table~\ref{table:sota} overviews the state of the approaches in computer vision for resource constrained devices. In the following, we describe each approach.

% Table~\ref{table:sota} presents an overview of approaches, centered around four major architectures: (1) On-device computation, where a ML/DL model runs on device. (2) On-edge computation, where a ML/DL model runs on edge device and (3) On-cloud computation, where a ML/DL model runs on cloud.  In the following, we describe each approach:

Table~\ref{table:sota} presents an overview of approaches, centered around three major architectures for video analytics on IoT devices~\cite{murshed2019machine, edge-dl-survey, deeplearning-edge, chauhan2016development, intizar:emse-01644333, 7460669, 10.1145/3041021.3054736, GYRARD2017305}: (1) \textbf{On-Device Computation}, where an ML or DL model runs on device (in Section~\ref{sec:classical-cv},~\ref{sec:cnn}). (2) \textbf{On-Edge Computation}, where an ML/DL model runs on powerful CPU or GPU Edge devices, instead of running on the device itself~(in Section~\ref{sec:on-edge}). (3) \textbf{Cloud-based Software as a Service~(SaaS)}, where an ML/DL model runs on cloud as a service~(in Section~\ref{sec:on-cloud}). In the following section, we describe each approach.

\begin{table*}[!ht]
\centering 
\scriptsize
\begin{tabular}{p{1.1cm}  p{6cm}  p{2.7cm} p{2.3cm}  p{4cm}}
 \toprule\textbf{Approaches}&\textbf{Description}&\textbf{Strengths}&\textbf{Weaknesses}&\textbf{Tools \& Technology} \\ \midrule 
\multirow{1}{*}{\parbox{1.3cm}{On-Device ML Techniques}} & \nextitem{Hand-engineered feature extractors \& classifiers to detect objects.} \nextitem{Model is trained on Edge or Cloud server.} \nextitem{Model inference on device.} & \parbox{1.6cm}{\nextitem Low Latency} 
\nextitem{No vendor lock-in}
\nextitem{Easy to understand} 
 & \nextitem{Accuracy} \nextitem{Maintenance} \nextitem {Development Effort}
 &  \parbox{3cm}{\nextitem{Haar Cascade}} \parbox{3.2cm}{\nextitem{HOG and SVM}} 
 \parbox{4cm}{\nextitem{SIFT}} 
 %\nextitem{Classical Computer Vision Algo. - OpenCV} \nextitem{Tensorflow}
 \\ \midrule 
%\multirow{1}{*}{\parbox{1.3cm}{WSN Database}} & It provides SQL interface to extract data from sensing devices. & TinyDB\cite{madden2005tinydb}, IrisNet\cite{gibbons2003irisnet} & Easy to use interface to extract data from a large number of devices. & \nextitem{Less flexible for introducing the application logic.} \nextitem{Largely targets for similar types of devices.} \\ \midrule

\multirow{1}{*}{\parbox{1.3cm}{On-Device DL Techniques}}& \nextitem{Model design or Model compression techniques.} \nextitem{Model training on Edge server or cloud.} \nextitem{On device inference.}  & \nextitem{Low latency} \nextitem{No vendor lock-in} \nextitem{Accuracy} & \nextitem{Maintenance} \nextitem{Development effort} \nextitem{Difficult to understand} &  \nextitem{MobileNet SSD} \nextitem{YOLO}
\\ \midrule
\multirow{1}{*}{\parbox{1.1cm}{On-Edge Computation}} 
& \nextitem{Offloading inference from a device to nearby Edge server.} 
	%\nextitem{A device sends a video stream to a nearby edge server and receives the corresponding results after server processing.}
	\nextitem{Model training and training at Edge server.}
	\nextitem{Co-processor/accelerator device, dedicated GPU boards}
& \nextitem{Reduced Latency} \nextitem{No Vendor lock-in}
& \nextitem{Resource management} \nextitem{Development Effort} 
 & \nextitem{Intel Movidius Neural Compute Stick 2}  \nextitem{Google Coral TPU USB Accelerator}  
 \nextitem{NVIDIA Jetson Nano}
 %\nextitem{Google Coral Development Board} \nextitem{YOLO} 
\\ \midrule

\multirow{1}{*}{\parbox{1.5cm}{Cloud-based SaaS}} & \nextitem{Cloud-based Software as a Service (SaaS) platform.}
% \nextitem{ A device sends a video stream over the network to a cloud-based platform, where all the processing takes place and detection results are sent back to device.}
 \nextitem{Model inferencing and training at Cloud Platform.}
 & \parbox{2.5cm}{\nextitem{Scalable platform}}
 \parbox{2.3cm}{\nextitem{Rapid Development}}
 \nextitem{Anywhere Resource Access}
 \nextitem{Anytime Resource Access}
 & \nextitem{High Latency} \nextitem{Subscription cost} \nextitem{Privacy concerns} \nextitem{Vendor lock-in} 
 & \nextitem{AWS Rekognition} \nextitem{Google Vision AI} \nextitem{Microsoft Azure Computer Vision}
\\ \bottomrule													 
\end{tabular}
\caption{Summary: State of the Art in Video Analytics for IoT Devices.} 
\label{table:sota} 
\end{table*}

% Summary: State-of-the-art Approaches in Video Analytics for IoT devices.

\subsection{On-Device ML-based Techniques}\label{sec:classical-cv}

When designing a model, the efforts belonging to this category often focus on designing a simple model that can run on IoT devices. 
The overall objective of this simple model is to reduce the execution latency, while aiming to maintain acceptable levels of accuracy. In the following, we present some of the On-Device ML-based techniques: 

%In the following section, we present some of the popular models in literature for IoT-based video analytics.

% The overall objective of this simple model is to reduce memory and execution latency, while aiming to preserve reasonably good accuracy.

\fakeparagraph{Haar Cascade} Haar cascade is an object-detection technique to detect objects, based on the concept proposed by Viola-Jones~\cite{haar}. It is an ML-based approach where a cascade function is trained using a set of positive and negative images. The first step of this technique is to collect Haar features from an image. A large number of Haar features are necessary to describe an object from an image, as a Haar feature is a weak classifier. Therefore, they are organized into cascade classifiers to form a strong classifier to detect an object from an image. We have implemented and trained a full custom object Haar cascade detection.  \textbf{The performance results of Haar cascade are presented in Section~\ref{sec:evaluation}}.

\fakeparagraph{HOG and SVM} \textbf{H}istogram of \textbf{O}riented \textbf{G}radients~(HOG)~\cite{hog} is a widely used feature descriptor to extract features from an image. An image is broken down into smaller regions. The gradient and orientations (or magnitude and directions) are calculated for each region. Then, HOG generates a histogram of each region separately. The gradient and orientations of pixel values are used to create the histograms, therefore this technique is named as “Histogram of Oriented Gradients (HOG)”. HOG features (called as feature vector) are given to Support Vector Machine (SVM)~\cite{svm} classifier for object detection. SVM is a supervised ML model that uses classification algorithms for binary classification. For our experiments, we have trained linear SVM that classifies object (e.g., human) from non-object (e.g., non-human). \textbf{We present the performance results of HOG + SVM in Section~\ref{sec:evaluation}.} 

The On-Device ML-based techniques are light-weight and show low execution latency, which makes these algorithms a suitable candidate for on-device video analytics. However, the real-world objects could have different lighting conditions and appearance, where it is harder for these types of models to achieve a high detection accuracy~\cite{from-haar-svm-cnn}. To achieve a high accuracy, DL-based techniques have been proposed, discussed in Section~\ref{sec:cnn}.

% When designing an object recognition model for resource constrained devices, the efforts belonging to this category often focus on designing a simple model of object detection, thus reducing memory and execution latency, while aiming to preserve reasonably good object recognition accuracy (not high as CNN-based models, as discussed in the following section).  Many classical computer vision algorithms such as Haar-Cascade, Histograms of Oriented Gradient~(HOG) with the support of Support Vector Machine~(SVM), and Scale Invariance Feature Transformation~(SIFT) can be used for object detection, as these models are light weighted and very fast. 

% \subsection{CNN-based model for resource constrained devices}\label{sec:cnn}
\subsection{On-Device DL-based Techniques}\label{sec:cnn}
The approaches, belonging to this category, employ various model reduction techniques to enable its deployment on IoT devices. For instance, \textbf{Model design techniques} emphasize on models design with a reduced number of parameters~(while aiming to maintain high levels of accuracy), thereby bringing down the memory requirements to run the model on IoT devices~\cite{deeplearning-edge}. \textbf{Model compression} techniques such as parameter quantization, parameter pruning have been proposed~\cite{deeplearning-edge} to reduce model size. In parameter quantization, the parameters of the existing DL model are compressed by going to low-bit width numbers from floating-point numbers. This takes out the expensive floating-point multiplications. In parameter pruning, parameters that are not so important~(like the ones which are near to zero) are removed. MobileNet SSD~(Single Shot Detector)~\cite{mobilenet-ssd, ssd} and YOLO~(You Only Look Once)~\cite{yolo} are some of the DL models for IoT devices for video analytics. Many of the DL models for IoT devices, with pre-trained weights, are available for download. For instance, Google’s TensorFlow Lite offers pre-trained and optimized models to identify hundreds of classes of objects, including people and animals.

% \textbf{Model design techniques} focus on designing models with a reduced number of parameters of a model, thus reducing memory and execution latency while aiming to preserve high accuracy~\cite{deeplearning-edge}.

% Parameter quantization takes an existing DL model and compresses its parameters by changing from floating-point to low-bit width numbers, thus avoiding costly floating-point multiplications. Parameter pruning involves removing the least important parameters such as parameters that are close to zero.

Many frameworks~(e.g., TensorFlow Lite, MXNet, Caffe2, etc.) have been proposed to deploy DL models on IoT devices~\cite{murshed2019machine}. For our experiments, we have used Google's TensorFlow Lite, which is a set of tools that help developers to run TensorFlow models on resource-constrained devices. It enables on-device machine learning inference with low latency and a small binary size. \textbf{We present the performance results of MobileNet SSD model for TensorFlow Lite in Section~\ref{sec:evaluation}.}

\subsection{On-Edge Computation}\label{sec:on-edge}
While the model reduction techniques can help DL models to run on resource constrained IoT devices, it is still a challenge to deploy a powerful and large DL model with a real-time execution. To address this challenge, this approach considers offloading DL model from an IoT device to more powerful entities or Edge device. In such scenario, an IoT device sends a video stream to a nearby Edge device, the Edge device processes the data and sends the corresponding results back to the IoT device. 

To speed up inference or training at Edge server, hardware manufacturers have developed CPUs and GPUs-enabled devices for deep learning. For instance, Intel's Movidius Neural Compute Stick 2~(NCS)~\cite{Intel2020} and Google's Coral Tensor Processing Unit~(TPU) USB Accelerator~\cite{Google2020} are some of the co-processors that can be plugged to the USB port of an Edge device. There are times when CPU-enabled devices with co-processor may not be enough for running DL models. To address this issue, hardware manufacturers offer GPU-based dedicated development boards, which tend to have good compatibility and performance. NVIDIA's Jetson Nano~\cite{nvidia2020} and Google Coral Development Board~\cite{googledev2020} are some of the examples of dedicated development boards to run DL models.

\subsection{Cloud-based Software as a Service~(SaaS)}\label{sec:on-cloud}
This approach leverages cloud-based SaaS for video analytics. Various cloud vendors~(Such as Amazon, Google, Microsoft) have developed DL-based platforms. A typical approach is where a device sends a video stream over the network to a cloud-based platform, where all the processing takes place and detection results are sent back to the IoT device.

For our implementation, we use AWS as a base technology. We choose AWS Rekognition service, which lets developers develop several computer vision capabilities on top of scalable and reliable Amazon infrastructure. AWS Rekognition offers services, which can be divided into two categories: First, the developer can leverage pre-trained algorithms (prepared by Amazon) to identify objects, people, text, scenes, and activities in videos, as well as detect any inappropriate content. Second, AWS Rekognition Custom labels enable the developers to build use case specific ML-based video analytics capabilities to detect unique objects and scenes. \textbf{We present the performance results of AWS Rekognition in Section~\ref{sec:evaluation}.}

\section{System Design and Implementation}\label{sec:sysdesign}

%Figure~\ref{fig:abc} illustrates an architecture of the proposed system. 

The proposed system consists of three layers based on their functions: \textbf{Device layer}, \textbf{Edge layer}, and \textbf{Cloud layer}. The data flow goes as follows: A real-time media data is first captured by the camera at the \textit{Device layer} and then it is transmitted to the \textit{Edge layer} for processing. The \textit{Edge layer} implements a video analytics logic. The video analytics results are sent to the \textit{Cloud layer} for storage and remote access. The Cloud layer exposes application APIs. The Cloud layer implements a MobileApp, WebApp and Voice interface~(on top of APIs exposed by the Cloud layer) that lets the users interact with the doorbell anywhere and anytime.

In the following sections, we describe the functionality of each layer and its implementation details.

\subsection{Device Layer}\label{sec:device-layer}
\begin{figure*}[h]
  \centering
  \includegraphics[width=\linewidth]{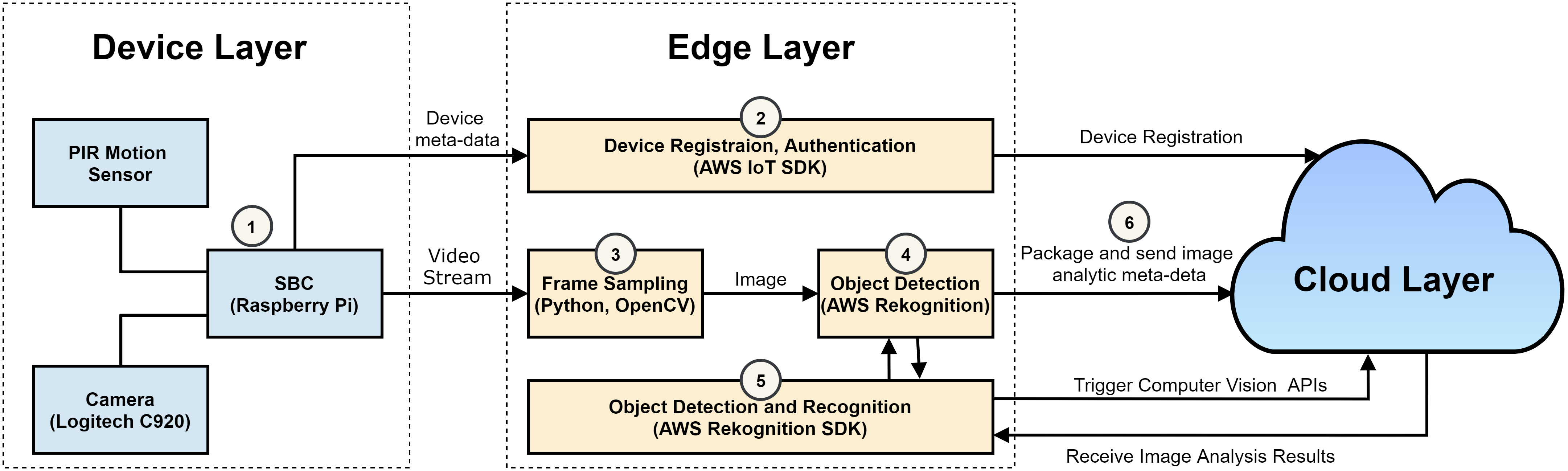}
  \caption{Logical Flow of Device and Edge Layer.}
  \label{fig:device-edge}
\end{figure*}

It contains an array of devices. Each device interfaces with a camera module to capture a video and has a PIR sensor to detect the motion of an object (Circled~\circled{1} in Figure~\ref{fig:device-edge}). The camera is used to capture video data. The motion sensor triggers the camera if any motion is detected. The integration of motion sensors with the device allows it to process data only when there is a motion. This would avoid the processing of redundant and uninteresting data. We prototype the device layer using WiFi-enabled Raspberry Pi 3 Model B+. 

% However, the device layer is flexible to integrate IP-camera as well. In the case of IP-camera, we set up a webserver in IP camera that listens for incoming requests and serves video footage to multiple clients simultaneously over Real-time Streaming Protocol~(RTSP).

Each device implements device registration and authentication functionality that allows the user to interact with the device anywhere and anytime~(Circled~\circled{2} in Figure~\ref{fig:device-edge}). We implement these functions using AWS IoT Core, which is a platform that enables the users to connect various types of devices to secure registration and authentication to AWS. More specifically, the implementation leverages two services of AWS IoT Core. First, device registry to keep a record of all the registered devices and their attributes. Second, the authentication service provides the mutual authentication service between the device and the Cloud, so that device data is never exchanged between the device and the Cloud without proven identity. It supports X.509 certificate-based authentication as well as certificates signed by a Certificate Authority~(CA).

Each device hosts an Edge component~(discussed in Section~\ref{sec:edge-layer}) to process data before the data is pushed to the Cloud.

% \begin{figure*}[h]
%   \centering
%   \includegraphics[width=\linewidth]{device-edge}
%   \caption{Logic Flow of Device and Edge Layer. \textcolor{red}{Redraw the image.}}
%   \label{fig:device-edge}
% \end{figure*}

\subsection{Edge Layer}\label{sec:edge-layer}
The Edge layer processes the transmitted media data, before the data is sent to the Cloud. Figure~\ref{fig:device-edge} presents a logical flow of these functions at the Edge and their interaction with the device and Cloud layer. In the following section, we present functions implemented at the Edge layer:

\fakeparagraph{Frame sampling} It samples a frame off of a live video stream from the attached camera~(triggered by a motion sensor at the device layer). It packages and sends raw footage to the object-detection component~(Circled~\circled{3} in Figure~\ref{fig:device-edge}). The frame sampling client is implemented using OpenCV~(Open Source Computer Vision) library in Python. 

% OpenCV is an open source computer vision and machine learning library.

\fakeparagraph{Object detection and recognition}  It implements a logic to detect objects. It is dedicated to running various models to detect various objects~(Circled~\circled{4} in Figure~\ref{fig:device-edge}). The current version implements four models: a face detection and recognition, an animal detection, an unsafe content detection~(such as violence, gun etc.) and a noteworthy vehicle detection such as a fire truck and courier service~(e.g., FedEx, DHL, USPS) van. 

The object-detection component implements functions that leverage AWS Rekognition APIs for object recognition and detection (Circled~\circled{5} in Figure~\ref{fig:device-edge}). AWS Rekognition is a highly scalable deep learning service that requires no machine learning expertise to use. It exposes various APIs to recognition objects. Table~\ref{table:aws-rekognition-apis} presents APIs used for the implementation.

\begin{table}[!ht]
\centering 
\footnotesize
\begin{tabular}{p{3cm}    p{4.7cm}   }
 \toprule
\textbf{AWS Rekognition APIs}&\textbf{Description} \\ \midrule 
 \texttt{search\_faces\_by\_image}  & It compares the features of the input face with faces in the specified database (or collection).  \\ \midrule
 \texttt{detect\_moderation\_labels}  & It detects unsafe content (such as gun, nudity, etc.) from a video.     \\ \midrule
 \texttt{detect\_text}  & It detects text in an input video and converts it into machine-readable text. We use this API to identify a noteworthy vehicle~(e.g., Ambulance, DHL truck) as  they embed text on top of the vehicle.   \\ \midrule
 \texttt{detect\_labels}  & It detects an instance of real-world entities~(such as animal, pet) within an input video stream.   \\  \bottomrule	
\end{tabular} 
\caption{AWS Rekognition APIs} 
\label{table:aws-rekognition-apis} 
\end{table}

The video analytics metadata~(e.g., object attributes such as known or unknown face, confidence, date and time, etc.) are sent to the Cloud for storage and further retrieval~(Circled~\circled{6} in Figure~\ref{fig:device-edge}), discussed in Section~\ref{sec:cloud-layer}.

\subsection{Cloud Layer -- Serverless Architecture }\label{sec:cloud-layer}

% \begin{figure*}[h]
%   \centering
%   \includegraphics[width=\linewidth]{cloud-layer}
%   \caption{Logic Flow of Cloud Layer. \textcolor{red}{Redraw the image.}}
%   \label{fig:cloud-layer}
% \end{figure*}

\begin{figure*}[h]
  \centering
  \includegraphics[width=\linewidth]{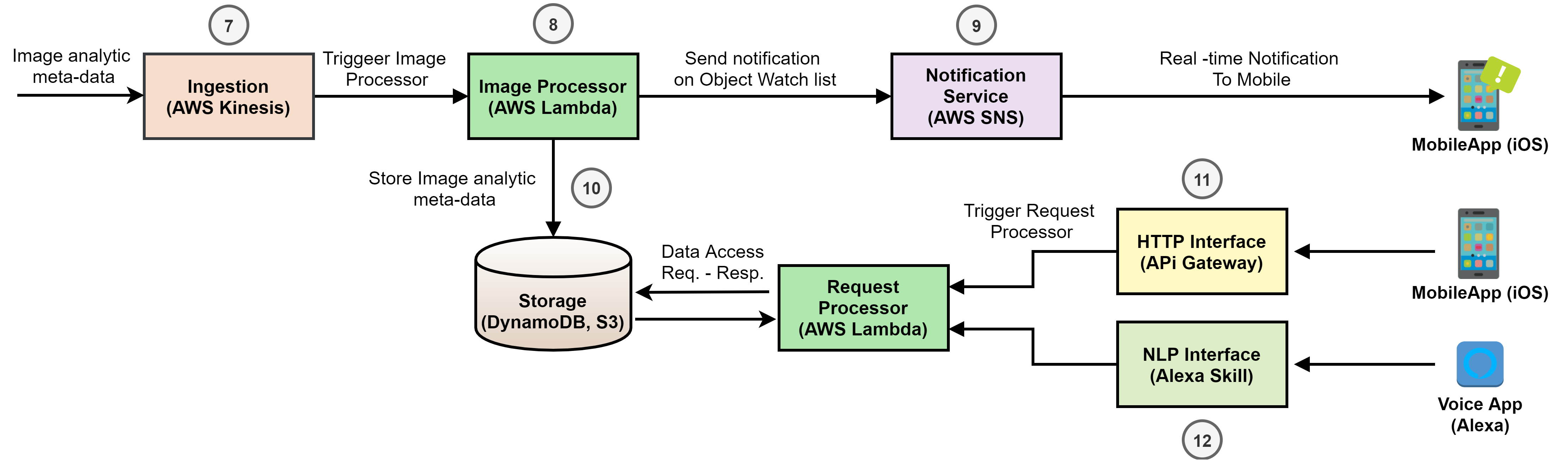}
  \caption{Logical Flow of Cloud Layer.}
  \label{fig:cloud-layer}
\end{figure*}

% The proposed Cloud layer implements serverless architecture or Function as a Service (FaaS), which is an agile solution for developers to build cloud-based services without heavy lifting of managing Cloud instances. We leverage AWS Lambda, which is an event-based and microservice framework in which a developer writes an application logic as lambda function(s). The application logic is triggered in response to the corresponding event. For example, a lambda function is triggered by an ingestion service~(Circled~\circled{8} in Figure~\ref{fig:cloud-layer}). AWS takes care of provisioning and resource management for executing lambda functions. We strongly advocate adopting serverless architecture, as the serverless architecture naturally solves two important problems: First, the serverless programming model greatly reduces the burden on developers in developing, deploying and managing applications, as there is no need to understand the complex underlying procedure to run applications or heavy lifting of distributed system management. Second, the functions are flexible to run on either Edge or Cloud with a minimum modification, which lowers the barrier of Edge-Cloud interoperability and federation. 

The proposed Cloud layer implements an architecture without a server~(Serverless) or in other words a Function as a Service~(FaaS).  It lets developers build Cloud-based services, while avoiding heavy lifting tasks of managing different instances of the Cloud.  We make use of an event-based and micro-service framework like AWS Lambda where a developer writes the application logic in the form of various function(s). The logic gets activated when a matching event occurs. For instance, an ingestion service triggers a lambda function~(Circled~\circled{8} in Figure~\ref{fig:cloud-layer}). AWS provisions and manages resources for the execution of functions of lambda. We strongly encourage using a serverless architecture because it inherently fixes two significant issues: First, programming without servers eases the work of developers to develop, deploy and manage applications to a great extent.  Second, the flexibility of functions for running  either on  Edge or Cloud with minor tweaks facilitates interoperability and federation between Edge and Cloud.

The Cloud layer is responsible for providing three essential functionalities: \textbf{scalable storage to access data anywhere and anytime},  \textbf{reliable processing infrastructure to  train new machine learning models}, and \textbf{API interfaces for user applications}. In the following section, we present each functionality:

\fakeparagraph{Model training and deployment} As the requirements evolve, the Cloud layer may need to build new models. For instance, a doorbell~(installed near a garage) would need a noteworthy vehicle detection model to recognize a car licence plate. To address this requirement, we integrate Amazon Rekognition Custom Labels, a recently launched (in November 2019) feature of Amazon Rekognition that enables the developers to build a customized (use case specific) ML-based video analytics model. With AWS Rekognition Custom Label, we can simply upload a small set of use case specific images~(a few hundred images, instead of requiring a larger number of hand-labeled images to build an accurate model) for training. In background, it automatically loads and inspects the training data and generates models, which can be exposed as Rekognition Custom APIs and integrated into our application.

%\footnote{\url{https://aws.amazon.com/rekognition/custom-labels-features/}}
%The proposed system is flexible to integrate new image analysis models on-fly.
% To address this requirement, the design should provide a reliable processing infrastructure to train new models. % Building an accurate model to analyze video or image often requires a larger number of hand-labeled images. This task may take months to gather required data and a large team of labelers to prepare it for use in machine learning.  Therefore, we choose Amazon Rekognition Custom Labels, a recently launched (in November 2019) feature of Amazon rekognition that enables the developers to build a customized (use case specific) ML-based image analysis model.

\fakeparagraph{Real-time Push Notification} It sends a real-time alert notification to the user, when a motion is detected in the proximity of the doorbell. This functionality is implemented using three components: AWS Kinesis, AWS Lambda, and Amazon SNS. We use AWS Kinesis Data Stream to ingest video analytics metadata from the Edge layer to the Cloud layer~(Circled~\circled{7} in Figure~\ref{fig:cloud-layer}). We implement AWS Lambda functions~(written using Python) that process the metadata in response to data ingestion from Kinesis~(Circled~\circled{8} in Figure~\ref{fig:cloud-layer}) and triggers the push notification~(Circled~\circled{9} in Figure~\ref{fig:cloud-layer}). We implement the push notification service using Amazon Simple Notification Service~(Amazon SNS), which is a web service that enables applications to instantly send notification from the Cloud.

\begin{figure*}[h]
  \centering
  \includegraphics[width=15cm, height=8.5cm]{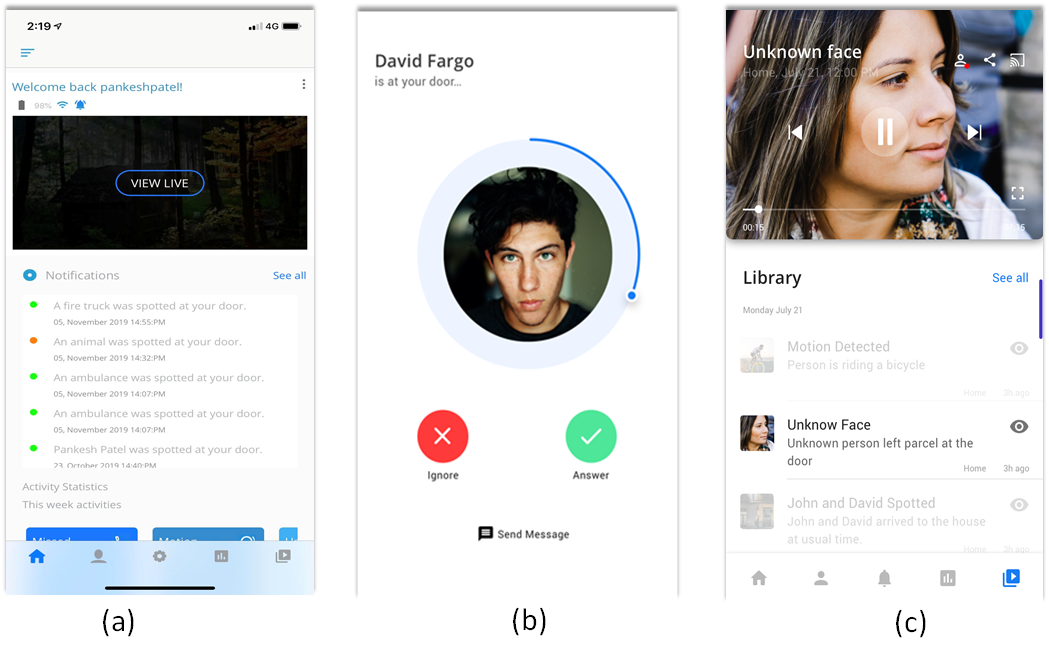}
  \caption{The user interface of the mobile application: \textbf{(a)} The lower part shows a dashboard that provides the detailed activities at the doorbell. The notification messages include a face detection of known and unknown persons and object detection (e.g., noteworthy car, animal, etc.), The upper part illustrates on-demand live video streaming that allows users to see what is happening at the doorbell. \textbf{(b)} Shows an interface for real-time push notification. The user receives alerts on his mobile application when a visitor is detected at the door. The user can respond to the notification or just "ignore" it. \textbf{(c)} Shows the video library. This interface lets users review activities and events at the door at a later time in case the user misses the real-time alert.
}
  \label{fig:mobile-app}
\end{figure*}

Figure~\ref{fig:mobile-app}~(b) shows an iOS mobile application interface of real-time push notification. The user receives alerts on his mobile application when a visitor is detected at the door. The user can respond to the notification or just "ignore" it.

% We implement an interface for real-time pish notification. The user receives alerts on his mobile application when a visitor is detected at the door. The user can respond to the notification or just ``ignore'' it. 

%\footnote{\url{https://aws.amazon.com/lambda/}}
%\footnote{\url{https://aws.amazon.com/sns/}}
%\footnote{\url{https://aws.amazon.com/kinesis/}}

\fakeparagraph{Persistent data storage and access} It receives video analytics metadata from Edge and provides a scalable storage to access data anywhere and anytime~(Circled~\circled{10} in Figure~\ref{fig:cloud-layer}). We implement the storage services using Amazon DynamoDB and Amazon S3. DynamoDB offers real-time database services, which store video analytics metadata in key-value format (Circled~\circled{10} in Figure~\ref{fig:cloud-layer}). Amazon offers Simple Storage Service (Amazon S3) to store and share user-generated content like images and video~(Circled~\circled{10} in Figure~\ref{fig:cloud-layer}). Both Amazon S3 APIs and DynamoDB APIs provide simple interfaces to store and retrieve data using the Amazon online storage infrastructure. We use these APIs that are exposed by Amazon API Gateway Service~(Circled~\circled{11} in Figure~\ref{fig:cloud-layer}). The API gateway is a service that sits in front of APIs. We use API gateway service because it allows us to encapsulate the storage APIs in mobile and web implementations. In our implementation, API gateway acts as a single point of entry and accommodates direct requests from MobileApp.

%\footnote{\url{https://aws.amazon.com/api-gateway/}}
%\footnote{\url{https://aws.amazon.com/dynamodb/}}
%\footnote{\url{https://aws.amazon.com/s3/}}

The lower part of Figure~\ref{fig:mobile-app}-(a) shows a dashboard that provides the detailed activities in the doorbell case study. The notification metadata includes a face detection of known and unknown persons and object detection~(e.g., noteworthy car, animal, etc.) which are stored in AWS DynamoDB. Figure~\ref{fig:mobile-app}-(c) shows the video library, stored in AWS S3. This interface lets users review activities and events at the door at a later time in case the user misses the real-time alert.

% We implement a dashboard that provides the detailed activities in the doorbell case study. The notification metadata includes a face detection of known and unknown persons and object detection~(e.g., noteworthy car, animal, etc.) which are stored in AWS DynamoDB. 
% We implement the video library, stored in AWS S3. This interface lets users review activities and events at the door at a later time in case the user misses the real-time alert.

\fakeparagraph{Conversational User Interface}
Our approach involves the voice assistant functionality, which allows the users to ask questions about the smart environment that they are in. The voice assistant system leverages the logged video analytics results to provide a meaningful response. We integrate a voice assistant approach for several reasons: first, the use of natural language has been considered an intuitive way for people to interact with the technology, specifically people who have limited mobility. Second, the deployment of voice assistants is becoming pervasive now in smart homes~(e.g., to date, over 100 million Amazon Echo devices have been sold~\cite{Matney2019}). 

To explore the feasibility and potential use of this approach, we integrate this approach into our Smart doorbell. We implement this conversation interface using Alexa voice service, which is an intelligent voice control service to power the devices such as Amazon Echo that is a smart personal assistant device from Amazon triggered using voice commands~(Circled~\circled{12} in Figure~\ref{fig:cloud-layer}). Alexa uses Natural Language Processing~(NLP) techniques to process users' requests. The voice service can be triggered using a specific keyword~(such as Alexa). We have implemented an Alexa skill that can be triggered using the various voice commands~(such as ``\textit{Alexa, tell me what is happening at the door?}'', ``\textit{Alexa, send me a snapshot of all activities at my door today}''). Our custom Alexa skill triggers a set of lambda functions in the Cloud, which queries DynamoDB, which stores the video analytics metadata. Once the query result is computed in the Cloud, the results are sent back to the user through Alexa voice service.

%\footnote{\url{https://developer.amazon.com/en-US/alexa/alexa-voice-service}}

\section{Evaluation}\label{sec:evaluation}
This section evaluates our AWS-Rekognition based proposed system design~(presented in Section~\ref{sec:sysdesign}) that supports the Smart doorbell use case, described in Section~\ref{sec:doorbell}. The evaluation results are divided into three different categories: The evaluation results of the proposed approach in Section~\ref{sec:model-accuracy}, the comparative analysis~(on Overall model performance, Latency, Memory and CPU usage) of the proposed approach with the state-of-the-art approaches in Section~\ref{sec:comparative-analysis} and Section~\ref{sec:cpu-memory-usage}.

\subsection{Model Accuracy}\label{sec:model-accuracy}

% We apply commonly used metrics for evaluating the performance of AWS Rekognition models for object detection and recognition.  Figure 1 presents our results of the scenarios, presented in the previous section. The results can be categorized in the following ways~\cite{abc}:

We apply commonly used metrics for evaluating the performance of AWS Rekognition-based proposed approach. We look at the following well-known metrics to derive model accuracy~\cite{Amazon2020}:

\fakeparagraph{True Positives~(TP)} The model correctly detects the presence of an object in an input image, when an object is present in the image. For instance, the model correctly returns an ambulance object, when an ambulance is present in an image.

\fakeparagraph{False Negatives~(FN)} The model does not detect an object that is present in an image, but the ground truth for the input image includes the object. For instance, the model does not return an "ambulance" for an image that contains an ambulance.

\fakeparagraph{False Positives~(FP)} The model incorrectly detects the presence of an object in an input image, when an object is not present in the image. For instance, the model detects an ambulance object, when there is no ambulance in the image.

\fakeparagraph{True Negatives~(TN)} The model correctly detects an object that is not present in an input image. For instance, the model does not detect an ambulance for an input image that does not contain an ambulance.

The above-mentioned metrics are often used to derive some well-known metrics to measure the predictive performance of a model: \textbf{Accuracy}, \textbf{Precision}, and \textbf{Recall}. Accuracy is the ratio of correct predictions~(i.e.,TPs) over the total number of predictions. Precision is the fraction of correct detection (i.e., TPs) over all positive results~(i.e., TPs and FPs). Recall is a measure of how often a model can predict an object in a video stream correctly when it is present in the video stream. Recall is the fraction of correct detection~(i.e.,TPs) over TPs and FNs.

\fakeparagraph{Datasets} We employ several datasets to evaluate the accuracy of the model. For face recognition scenario, we have used Labeled Faces in the Wild~(LFW), which is a database of face photographs designed for studying face recognition. This dataset contains more than 13,000 images of faces collected from several sources from the Web. Each face has been labeled with the name of the person's picture. This would help us to perform known or unknown face detection experiments. For the remaining scenarios, the dataset is self-collected from the various sources from the Web. The images have been collected from other datasets such as Pascal VOC 2012, Common Objects in Context~(COCO), ImageNet, and public sources~(such as Kaggle and Google images), which have been checked by humans. Some of these images represent challenging scenarios for models to detect objects, as the visible features vary when the distances and angles are different, but sometimes an object is partially visible or in different gestures. Many algorithms may not identify an object or incur many FP rates.

%\footnote{\url{https://www.kaggle.com/atulanandjha/lfwpeople}}

% The above mentioned metrics are often used to calculate precision and recall~\cite{abc}. Precision is the fraction of correct detections (true positives) over all model detections (true and false positives). Recall is a measure of how often the model can detect an object correctly when it is present in the images of input. 

% \begin{table}[!ht]
% \centering 
% \footnotesize
% \begin{tabular}{p{3.8cm}    p{4.2cm}   }
%  \toprule
% \textbf{AWS Rekognition APIs}&\textbf{Description} \\ \midrule 
%  \texttt{search\_faces\_by\_image}  & It compares the features of the input face with faces in the specified database (or collection).  \\ \midrule
%  \texttt{detect\_moderation\_labels}  & It detects unsafe content (such as gun, nudity, etc.) in an image.     \\ \midrule
%  \texttt{detect\_text}  & It detects text in an input image and converts it into machine-readable text. We use this API to identify a noteworthy vehicle~(e.g., Ambulance, DHL truck) as  they embed text on top of the vehicle.   \\ \midrule
%  \texttt{detect\_labels}  & It detects an instance of real-world entities~(such as animal, pet) within an input image.   \\  \bottomrule	
% \end{tabular} 
% \caption{AWS Rekognition Model Accuracy, Precision and Recall. \textcolor{red}{Re-draw and replace the table with our google doc.}} 
% \label{table:experiment-setup11} 
% \end{table}

Table~\ref{table:aws-model-accuracy} summarizes the performance results of AWS Rekognition model for our scenarios. The model outperforms in terms of precision because the model does not incorrectly detect the presence of an object in an input image. AWS Rekognition has accuracy above 90\% in all scenarios. The high values of TPs and TNs and the low value of FNs indicate the high accuracy of the model. AWS Rekognition has recall value above 80\% in all scenarios. This is because FNs is between 2\% to 10\% where the confidence value of an object detection is 90. Reducing it to 70 eliminates all FNs in our dataset while improving overall object recognition success rate. 

\begin{table}[!ht]
\centering 
\footnotesize
\begin{tabular}{p{3cm} p{0.1cm} p{0.1cm} p{0.1cm} p{0.3cm} p{0.7cm}  p{0.7cm} p{0.7cm }} 
 \toprule
\textbf{Scenario}& \textbf{TP} & \textbf{FN} & \textbf{FP} & \textbf{TN}   &\textbf{Accuracy (In \%)}&\textbf{Precision (In \%)}&\textbf{Recall (In \%)} \\ \midrule 
  Face Recognition & 45\% & 5\% & 0 & 50\% & 95 & 100 & 90  \\ \midrule
  Unsafe Content Detection & 44\% & 6\% & 0 & 50\% & 94 & 100 & 88 \\ \midrule
  Animal Detection & 40\% & 10\% & 0 & 50\% & 90 & 100 & 80  \\ \midrule
  Noteworthy Vehicle Detection & 43\% & 7\% & 0 & 50\% & 93 & 100 & 86  \\ \midrule
  Multiple Objects Detection& 15\% & 2\% & 0 & 83\% & 98 & 100 & 88   \\  \bottomrule	
\end{tabular} 
\caption{AWS Rekognition-based proposed approach results.} 
\label{table:aws-model-accuracy} 
\end{table}

\subsection{Comparative Analysis}\label{sec:comparative-analysis}
This section presents the comparative analysis~(on Overall Model Performance and Latency) of the AWS Rekognition-based proposed approach with respect to the state-of-the-art approaches, presented in Section~\ref{sec:sota}.

\begin{figure}[h]
  \centering
  \includegraphics[width=\linewidth]{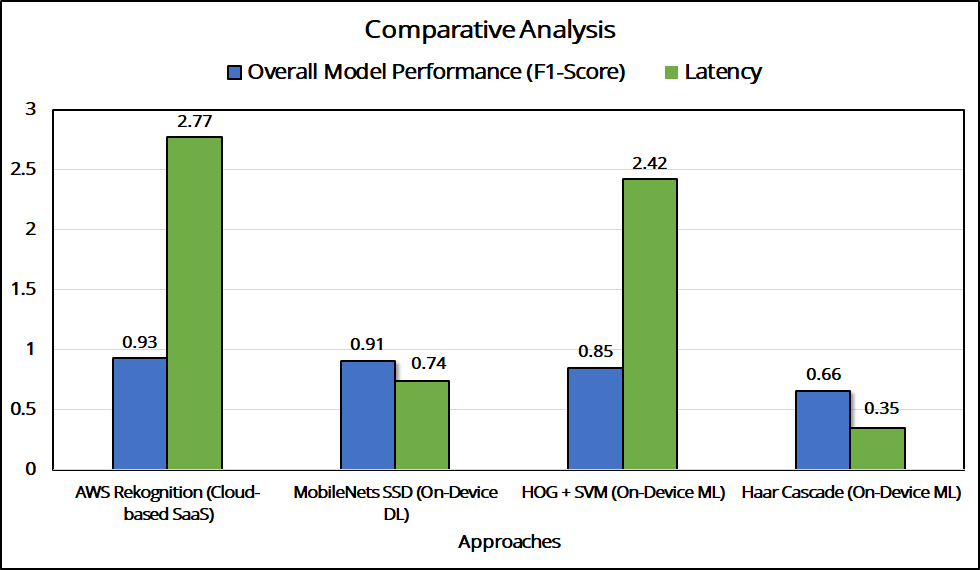}
  \caption{Comparative Analysis Results: Overall Model Performance and Latency.}
  \label{fig:f1-score}
\end{figure}

\fakeparagraph{Overall Model Performance} It is an aggregate measure that considers both recall and precision metrics of a model~\cite{Amazon2020}. We take the F1-score to measure the overall performance of a model. The F1-sore is a harmonic mean of precision and recall metrics. The value is between 0 and 1. The higher the value, the better the model performs, both on precision and recall. Figure~\ref{fig:f1-score} compares different approaches with the proposed AWS Rekognition-based model, including well-known models such as MobileNet SSD~(implemented using TensorFlow Lite framework), Haar Cascade, and HOG + SVM. The evaluation shows that AWS Rekognition has the highest F1-score among the compared approaches, indicating that the model is performing well both on precision and recall. The DL-based MobileNet SSD is close to the AWS Rekognition model.

% The evaluation shows AWS Rekognition and DL-based Tensorflow lite model almost have the same score, indicating that both the models have similar overall performance on the datasets we have used. However, AWS Rekognition has high latency compared to tensorflowlite model, discussed in the next section. 

%\subsection{Comparative Analysis on Latency}\label{sec:latency}

\fakeparagraph{Latency}
 It is the length of time it takes for a signal (or data) to be sent to the server plus the length of time it takes from the server for that signal to be received~\cite{dcomer}. Figure~\ref{fig:f1-score} summarizes the latency results of the proposed approach with respect to the state-of-art approaches. The result matches our intuition very well, showing that AWS Rekognition model exhibits high latency with respect to On-Device ML and On-Device DL approaches. The primary reason of having high latency is because of the communication between AWS Cloud and the doorbell over a network, that is to transfer data to AWS Cloud for video analytics and sending the outcomes~(or results) back to the doorbell. All the experiments are performed under strong wireless connectivity with 802.11n wireless LAN, although poor internet connection can dwarf these results and could increase the latency significantly.

\subsection{Memory and CPU usage}\label{sec:cpu-memory-usage}

We capture the memory utilization and CPU usage by our algorithms by a memory-profile module\footnote{\url{https://pypi.org/project/memory-profiler/}}, which is a Python module for monitoring memory and CPU usage of a process. This module gets the memory consumption of a process by querying the OS kernel about the amount of memory the current process has allocated.

Table~\ref{table:aws-memory-cpu} compares different approaches with the proposed AWS Rekognition-based algorithm, including Haar Cascade, HOG + SVM, and MobileNet SSD. The result matches our intuition very well in that DL-based algorithms require more memory and CPU because of its CNN-based Deep Neural Networks~(DNN), which may not be a good choice for highly resource-constrained IoT devices as they require 15 times more memory space. On the other hand, AWS Rekognition-based algorithm consumes very less memory because it offloads the video processing tasks on Cloud. This characteristic makes it suitable for resource-constrained IoT devices.

\begin{table}[!ht]
\centering 
\footnotesize
\begin{tabular}{p{3.2cm} p{1.9cm} p{1.9cm} } 
 \toprule
\textbf{Approaches}& \textbf{Memory Usage (in MBs)} & \textbf{CPU Usage~(in \%)}  \\ \midrule 
  AWS-Rekognition-based Approach (Cloud-based SaaS)
 & 1.99 & 28\%
  \\ \midrule
  MobileNet SSD~(On-Device DL)
 & 473.96
 & 33.30\%
  \\ \midrule
  HOG + SVM~(On-Device ML)
 & 30.09
 & 30\%
   \\ \midrule
  Haar Cascade~(On-Device ML)
 & 22.86
 & 30.20\%
   \\ \bottomrule	
\end{tabular} 
\caption{Approaches: Memory and CPU Usage} 
\label{table:aws-memory-cpu} 
\end{table}

% CPU usage refers to how much the processor is working. A device's CPU usage can vary depending on the types of tasks that are being performed by the processor.In our case the average CPU usage for the execution of the model is 28\%, with a maximum usage of 40\% while capturing frames from video.

% Memory usage refers to the sum total of the memory of the device that is utilized by the function on which it is being executed.

% More specifically for our scenario, the memory utilization is defined as the amount of memory of the edge device that is being occupied/engaged/consumed throughout the complete process of 
% detection and recognition of the objects.

% The results derived about the memory usage of the model states that while the detection and recognition function consumes only about 200 Kb of memory(on  average), other tasks such as 
% capturing  frames from a video and converting the captured frames back to video are the ones with maximum memory consumption  of 8388.61 Kb due to the storing of images and 7593.9839 Kb due to the mp4 video generated and stored in memory, which is followed by the task of converting the captured frame to byte array utilizing 925.696 Kb of memory and uploading the results to the AWS cloud utilizing 507.904 Kb of the memory  of the edge device. 

\section{Conclusion and Future Work}\label{sec:conclusion}
In this paper, we propose a distributed framework to orchestrate video analytics across Edge and Cloud resources. The proposed framework aims at the smart doorbell case study and uses AWS as a base platform for the implementation. To showcase the composability and reusability of the proposed framework, the ML-and DL-based models which are state of the art are implemented on the affordable Raspberry Pi in the form of an Edge device. This also demonstrates the feasibility of running these models on resource-constrained devices. Moreover, this paper evaluates the proposed framework as well as the state-of-the-art models and presents comparative analysis of them on various metrics~(such as overall model accuracy, latency, memory and CPU usage). The evaluation results demonstrate our intuition very well, showcasing that AWS-based approach exhibits high accuracy~(more than 90\% in all scenarios), low memory and CPU usage with respect to the state-of-the-art approaches, but high latency. These results show that AWS-based approach is quite suitable for applications such as Smart doorbell. However, it may not be suitable for latency-sensitive applications, because the delay caused by transferring data to Cloud for video analysis and back to the doorbell device is high, as supported by our evaluation results.

Our next objective is to reduce our reliance on Cloud-based SaaS platform to improve the existing latency results and reduce the subscription cost of a Cloud platform. 
Secondly, privacy is a major concern in the video-based smart doorbell application, as there is an involvement of bio-metric data of homeowners. Therefore, in the future, we plan to design a distributed Edge-based framework for low-cost resource-constrained IoT devices for video analytics applications. We potentially see Distributed Deep Learning and Federated Learning as solutions to address these challenges. 

\begin{acks}
This publication has emanated from research supported in part by a research grant from Science Foundation Ireland under Grant Numbers 16/RC/3918~(Confirm SFI Research Centre for Smart Manufacturing) and 12/RC/2289\_2~(Insight SFI Research Centre for Data Analytics), co-funded by the European Regional Development Fund. We thank Dr. Mohendra Roy and Dr. Pawan Sharma for their valuable inputs during  brainstorming sessions.
\end{acks}

\bibliographystyle{ACM-Reference-Format}
\bibliography{sample-base}

\end{document}